\documentclass[floatfix,reprint,amsmath,amssymb,superscriptaddress,aps,prb]{revtex4-2}
\usepackage{graphicx}  
\usepackage{dcolumn}   
\usepackage{bm}        
\usepackage{longtable}
\usepackage{amsmath}
\usepackage{ltablex}
\usepackage{color}
\usepackage{epstopdf}
\usepackage{adjustbox}
\usepackage{soul}
\usepackage{amsfonts}
\usepackage{tabularx}
\usepackage{text
comp, gensymb}
\usepackage{multirow}
\usepackage{array,booktabs}
\usepackage{hyperref}
\hypersetup{colorlinks=true, linkcolor=blue, citecolor=blue, urlcolor=black, pdftitle={Engineering topologically nontrivial structures by stacking topologically trivial materials} pdfauthor={xxx}}
\begin{document}

\title{Engineering stacking-induced topological phase transitions in bilayer heterostructures}

\author{Arjyama Bordoloi}
\affiliation{Department of Mechanical Engineering, University of Rochester, Rochester, New York 14627, USA}

\author{Daniel Kaplan}
\email{d.kaplan1@rutgers.edu}
\affiliation{Center for Materials Theory, Department of Physics and Astronomy, Rutgers University, Piscataway, New Jersey 08854-8019, USA}

\author{Sobhit Singh}
\email{s.singh@rochester.edu}
\affiliation{Department of Mechanical Engineering, University of Rochester, Rochester, New York 14627, USA}
\affiliation{Materials Science Program, University of Rochester, Rochester, New York 14627, USA}
\affiliation{Center for Coherence and Quantum Optics, University of Rochester, Rochester, NY, 14627, USA}

\date{\today}

\begin{abstract}
Nonmagnetic topological insulators (TIs) are known for their robust metallic surface/edge states that are protected by time-reversal symmetry, making them promising candidates for next-generation spintronic and nanoelectronic devices. 
Traditional approaches to realizing TIs have focused on inducing band inversion via strong spin-orbit coupling (SOC), yet many materials with substantial SOC often remain topologically trivial. 
In this work, we present a materials-design strategy for 
engineering topologically non-trivial phases, e.g., quantum spin Hall phases, by vertically stacking topologically trivial Rashba monolayers in an inverted fashion. 
Using BiSb as a prototype system, we demonstrate that while the BiSb monolayer is topologically trivial (despite having significant SOC), an inverted BiSb-SbBi bilayer configuration realizes a non-trivial topological phase with enhanced spin Hall conductivity. 
We further reveal a delicate interplay between the SOC strength and the interlayer electron tunneling that governs the emergence of a nontrivial topological phase in the bilayer heterostructure. 
This phase can be systematically tuned using an external electric field, providing an experimentally accessible means of controlling the system’s topology.
Our magnetotransport studies further validate this interplay, by revealing $g$-factor suppression and the emergence a zeroth Landau level. 
Notably, the inverted bilayer heterostructure exhibits a robust and tunable spin Hall effect, with performance comparable to that of state-of-the-art materials. 
Thus, our findings unveil an alternative pathway for designing and engineering functional properties in 2D topological systems using topologically trivial constituent monolayers.

\end{abstract}


\maketitle


\section{Introduction}
The emergence of topological insulators (TIs) has highlighted the profound role of topology in condensed matter systems.
Nonmagnetic TIs exhibit exotic transport phenomena, such as the quantum spin Hall (QSH) effect, which features counter-propagating helical edge states with opposite spin~\cite{bernevig2006_science, Bernevig_PRL2006, qi2011_rev_mod_phy, QiZhang2010}, resulting in quantized spin Hall conductivity.
Protected by time-reversal symmetry, these helical edge states prevent electron backscattering, enabling dissipationless edge transport. 
The robustness of these edge currents against geometric variations and disorder makes TIs promising candidates for low-power nanoelectronic and spintronic applications~\cite{Tang2017_nature}. 
Driven by these intriguing properties, significant progress has been made over the past two decades in predicting and designing two-dimensional topologically nontrivial systems (2D TIs)~\cite{konig2007_science,liu2008_PRL, konig2008, QiZhang2010, Maciejko2010, liu2011PRL, Tang2017_nature, knez2011_PRL,xu2013_PRL,weng2014_PRx,qian2014_science,zhu2015_nat_mat}, starting from their seminal discovery in graphene~\cite{kane2005_PRL, kane2005_PRL_2}.

Initial research efforts on 2D TIs  focused on identifying systems with intrinsic topological properties~\cite{Hasan_RevModPhys_2010, QiZhang2010, qi2011_rev_mod_phy}, but the field has since expanded to include strategies for inducing topological phases through engineered material architectures~\cite{Das2013_nature, Nechaev2017_sci_rep, Gunnink_2020_JPCM}.

Das \textit{et al.}~\cite{Das2013_nature} proposed a strategy for designing artificial 3D TIs by stacking Bi\textsubscript{2}Se\textsubscript{3} bilayers, each bilayer comprising two-dimensional Fermi gases with opposite Rashba spin-orbit coupling (RSOC)~\cite{Rashba1959, Rashba1960properties, bychkov1984oscillatory, bychkov_JETP_1984properties} on adjacent layers. They found that a nontrivial topological phase emerged after stacking three bilayers, while a massless Dirac point appeared only after six layers. 
An oscillatory crossover between trivial and topological phases in $\textrm{Bi}_2\textrm{Se}_3/\textrm{Te}_3$ was also predicted by Liu and coworkers~\cite{Liu2010oscillatory} and experimentally tested in Ref.~\cite{Kim2011}. Evidence for thickness-dependent topology in ZrTe\textsubscript{5} was reported in Ref.~\cite{lu2017thickness}. 
Building on these advances, Nechaev \textit{et al.}~\cite{Nechaev2017_sci_rep} designed a 2D QSH phase using a centrosymmetric sextuple layer, formed by stacking two BiTeI trilayers with opposite RSOC. Similar studies have also been conducted on heterostructures involving transition metal oxides with heavy ions~\cite{Gunnink_2020_JPCM}.

Previous efforts to engineer artificial TIs have largely focused on multi-layer configurations, exploring how topological properties evolve with the number of vertically stacked bilayers~\cite{Das2013_nature, Gunnink_2020_JPCM, Nechaev2017_sci_rep}. 
However, a significant gap remains in understanding the specific impact of the van der Waals (vdW) nature of the layers -- particularly its role in mediating interlayer electron tunneling and electronic band hybridization -- on the topological properties of layered vdW materials.
Moreover, much of the existing research has focused exclusively on the structural design of TIs, with limited exploration of application-oriented aspects, such as the influence of external electric and magnetic fields on spintronic functionalities~\cite{Das2013_nature, Gunnink_2020_JPCM, Nechaev2017_sci_rep}. 
These gaps highlight the need for innovative approaches that integrate material architecture with functional tunability in the design of 2D topological quantum phases for practical applications.

\begin{figure*}[tbh]
\centering
\includegraphics [width=17cm]{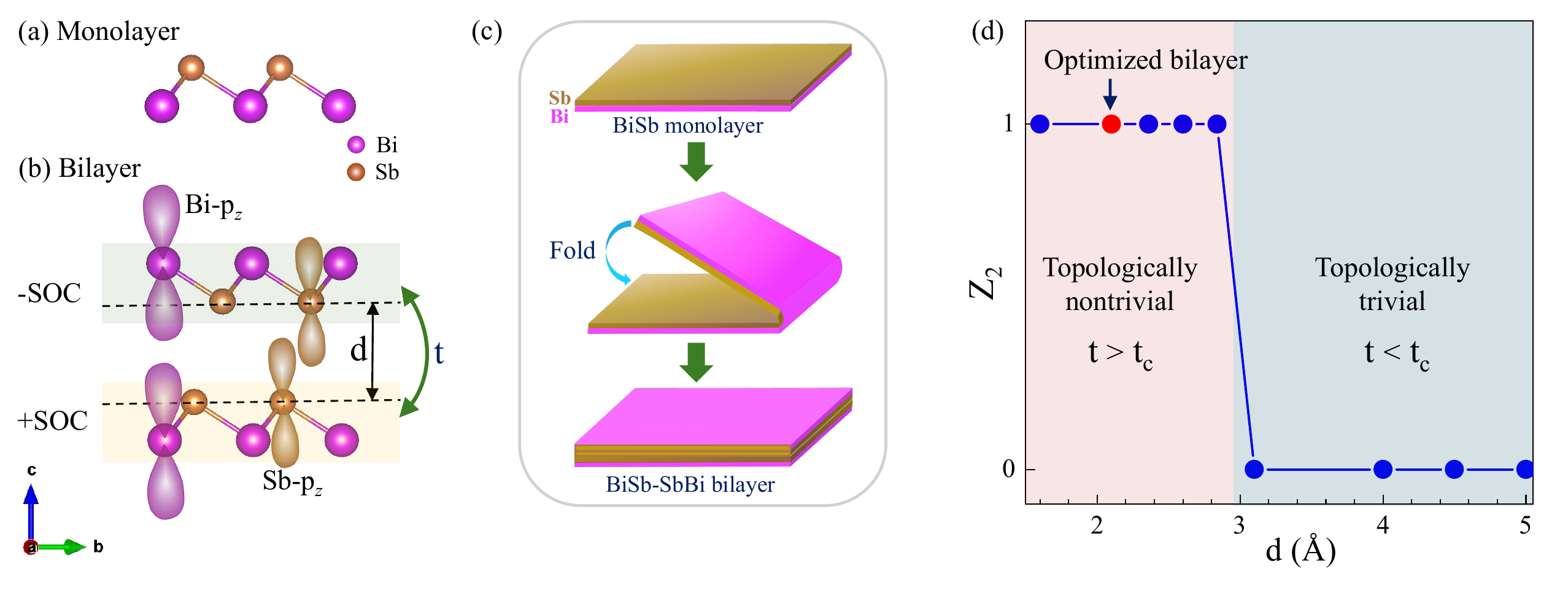}
\caption{Optimized crystal structures of (a) BiSb monolayer and (b) BiSb-SbBi bilayer, where \(d\) represents the interlayer distance between the two inverted monolayers and \(t\) denotes the interlayer electron tunneling. (c) Schematic illustration of the proposed experimental design scheme for realizing the BiSb–SbBi bilayer structure by folding a BiSb monolayer. (d) Variation of the Z$_2$ topological invariant in the BiSb-SbBi bilayer with varying interlayer distance \(d\). The light red region indicates the topologically nontrivial phase with Z$_2$\,=\,1, while the light blue region represents the topologically trivial phase with Z$_2$\,=\,0.}
\label{fig:crystal structure}
\end{figure*}

In this work, we demonstrate an alternative approach to engineering a 2D topologically nontrivial phase by vertically stacking two topologically trivial BiSb monolayers ($Z_2 = 0$) in an inverted configuration.
The resulting BiSb-SbBi bilayer exhibits a nontrivial topological phase ($Z_2 = 1$),
which is strongly sensitive to a delicate interplay between the SOC strength and the interlayer electron tunneling. 
Our first-principles density-functional theory (DFT) calculations reveal that this topological behavior can be tuned via interlayer spacing, SOC strength, and/or an external electric field, enabling reversible control over the system’s topological state. 
By carrying out magnetotransport studies, we show that the topologically nontrivial phase exhibits suppressed 
spin splitting in Landau levels, leading to a reduced 
effective $g$-factor, in contrast to the trivial phase. These interesting magnetotransport features, resulting from the interplay between interlayer electron tunneling and strong SOC in the individual layers~\cite{Kaplan2024}, could serve as a distinctive experimental signature for topological phase transitions, as has recently been observed in $\textrm{Bi}_2\textrm{O}_2\textrm{Se}$~\cite{wang2024even}. 
The bilayer also exhibits substantial spin Hall conductivity (SHC) [$\sim$178\,($\hbar$/e)\,S/cm] at the Fermi level, surpassing many 2D TIs~\cite{Zuo_PhysRevB_2023,Li_PhysRevB_2021} and rivaling that of known bulk TIs~\cite{Farzaneh_PhysRevMaterials.4.114202}. 

Given that BiSb thin films have already been experimentally synthesized~\cite{Lorenz_Thin_Solid_Films_2022, Fan__Japaneese_JAP_2020}, our findings not only establish a viable strategy for realizing tunable topological phases in layered vdW materials, but also highlight BiSb bilayers as a realistic and experimentally accessible platform for advancing spintronic and quantum technologies.

\section{Computational details}
\label{sec:comp_details}
First-principles density functional theory (DFT) calculations \cite{HK_dft_1964, KS_dft_1965} were performed employing the projector augmented wave (PAW) method \cite{Blochl94} as implemented in the Vienna Ab initio Simulation Package (VASP) \cite{Kresse96a, Kresse96b, KressePAW}. The kinetic energy cutoff for the plane wave basis set was set to 650 eV, and the exchange-correlation part of the Hamiltonian was computed using the generalized-gradient approximation as parameterized by Perdew, Burke, and Ernzerhof for solids (PBEsol) \cite{PBEsol}. Energy convergence criterion for electronic self-consistent calculations was set to ${10}^{-7}$ eV. Lattice parameters and inner-atomic coordinates were fully optimized until residual Hellmann-Feynman forces were less than ${10}^{-3}$ eV/Å per atom. The Brillouin zone was sampled using a $\Gamma$-centered $k$-mesh of size 12\,$\times$\,12\,$\times$\,1. In the PAW pseudopotential, contributions from five electrons each from Bi ($6s^26p^3$) and Sb ($5s^25p^3$) were considered. When constructing monolayer and bilayer configurations, a vacuum thickness of more than 15\,Å was maintained, and van der Waals corrections were considered using the DFT-D3 approximation \cite{Grimme_D3}. VASPKIT~\cite{VASPKIT} and the {\sc PyProcar} package~\cite{pyprocar} were employed for post-processing of the electronic structure data.  Dynamical stability was investigated through phonon calculations using 4\,$\times$\,4\,$\times$\,1 supercells within the finite displacement method as implemented in VASP, with postprocessing done using PHONOPY \cite{phonopy}. Topological properties were computed using Wannier tight-binding Hamiltonians with projections from s, $p_x$, $p_y$, and $p_z$ orbitals of Bi and Sb atoms computed using Wannier90 \cite{Marzari2012}. Subsequent calculations of Z$_2$ topological invariants and SHC from the Wannier tight-binding Hamiltonian were done using WannierTools \cite{WU2018405}. For computing the SHC, a dense k-mesh of 251\,$\times$\,251\,$\times$\,1 was used, ensuring convergence of SHC values within less than 1\% by varying the k-mesh size.

\section{Results and Discussions}
\subsection*{Crystal structure and dynamical stability}
The free-standing BiSb monolayer crystallizes into $p3m1$ layer group (no.\,\#$\,69$), as shown in Fig.~\ref{fig:crystal structure}(a). In this work, we design a bilayer configuration (BiSb-SbBi) by stacking two BiSb monolayers on top of each other in an inverted fashion, as depicted in Fig.~\ref{fig:crystal structure}(b). This ensures that the two constituent monolayers possess opposite RSOC. 
Unlike the monolayer, the bilayer restores space-inversion symmetry resulting in $p\bar{3}m1$ layer group (no.\,\#\,72). The DFT-optimized lattice parameters are $a\,= b\,= 4.169 \, \text{\AA}$, with an equilibrium interlayer vdW distance ($d_0$) of $2.1 \, \text{\AA}$.
Our DFT calculations confirm the dynamical stability of both the BiSb monolayer and the BiSb-SbBi bilayer (Fig.\,\ref{phonon}),
indicating their feasibility for experimental realization.

Given that BiSb thin films have already been synthesized~\cite{Lorenz_Thin_Solid_Films_2022, Fan__Japaneese_JAP_2020}, BiSb–SbBi bilayers could potentially be fabricated experimentally either by folding a single BiSb monolayer, as schematically illustrated in Fig.~\ref{fig:crystal structure}(c), or through state-of-the-art  thin-film deposition techniques.
The simulated Raman spectra for both the monolayer and bilayer are provided in the Appendix~\ref{appendix B} (Fig.\,\ref{Raman}), offering distinct fingerprints to aid future experimental characterization.

\subsection*{Electronic structure and tunable topological properties}

The free-standing BiSb monolayer is a topologically trivial Rashba semiconductor~\cite{Singh_PhysRevB_2017}. 
However, when two BiSb monolayers are stacked in an inverted configuration, as illustrated in Fig.~\ref{fig:crystal structure}(b), a 2D TI phase emerges in the bilayer configuration. 
The topological properties of the configurations are analyzed by computing the Z$_2$ topological invariant using the concept of Wannier charge centers (WCC)~\cite{Soluyanov_PhysRevB.83.035108}. A Z$_2$ invariant of 1 indicates a topologically nontrivial phase, whereas Z$_2 = 0$ corresponds to a trivial phase~\cite{kane2005_PRL}. 

\begin{figure*}
\centering
\includegraphics [width=18 cm]{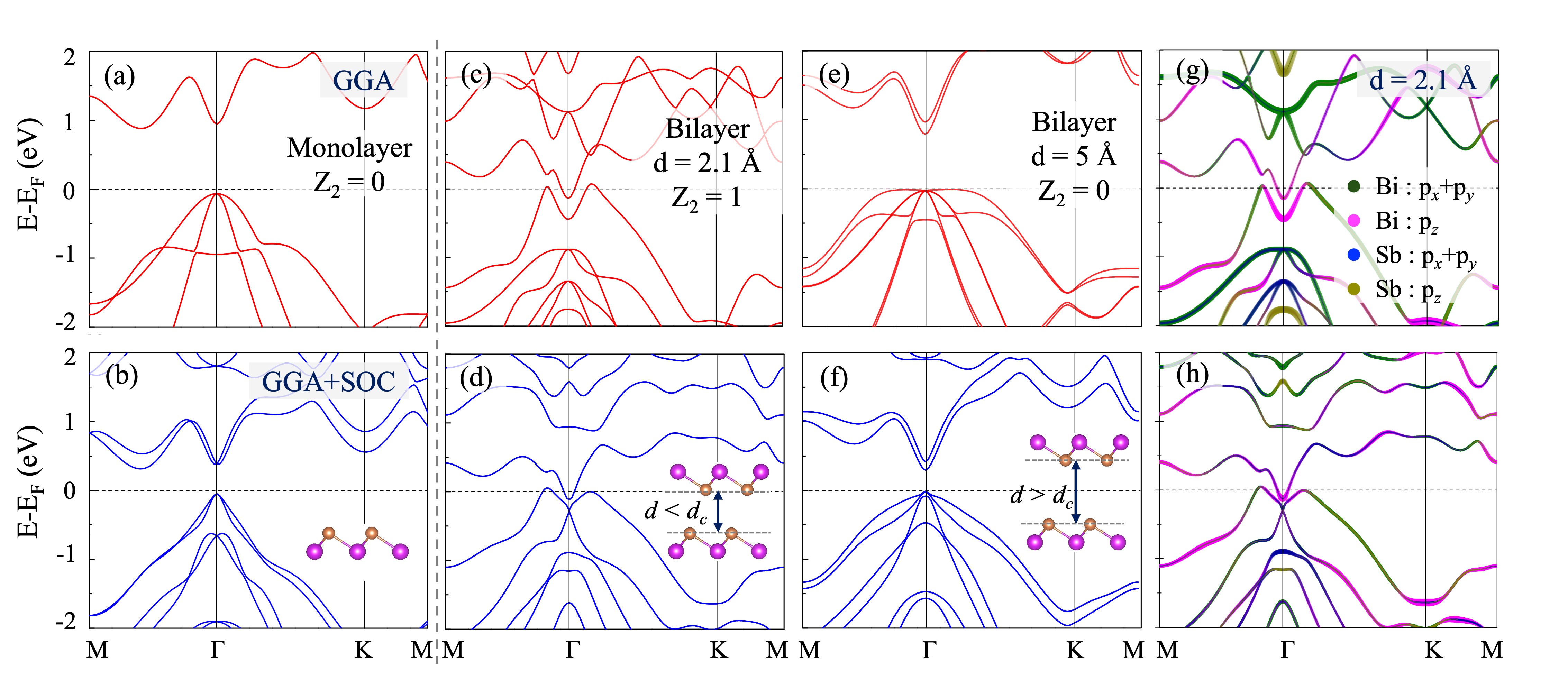}
\caption{Electronic band structure of the BiSb-SbBi bilayer with varying interlayer distance~($d$). Electronic band structures are computed along high-symmetry directions of the Brillouin zone for (a)-(b) the BiSb monolayer, (c)-(d) the bilayer at d = 2.1 Å (optimized), (e)-(f) the bilayer at d = 5 Å, and (g)-(h) the orbital-projected band structure at d = 2.1 Å (optimized). Horizontal dashed lines mark the Fermi level. The top panels show band structures without SOC (red), while the bottom panels include SOC (blue).}
\label{fig:band structure}
\end{figure*}

As illustrated in Fig.~\ref{fig:crystal structure}(d), the topological character of the bilayer is strongly dependent on the interlayer spacing\,($d$) between the monolayers. The optimized bilayer structure, with an equilibrium interlayer spacing $d_0 = 2.1 $\,\text{\AA}\, exhibits a nontrivial topological phase, which remains nontrivial as the layers are brought closer together. In contrast, increasing \(d\) beyond a critical value \(d_c = 2.84 \, \text{\AA}\) drives a transition to a topologically trivial phase, which persists at larger separations. To elucidate the underlying mechanism of this topological phase transition, we analyze the electronic band structures of both the monolayer and the bilayer at different values of \(d\) as shown in Fig.~\ref{fig:band structure}.

Fig.~\ref{fig:band structure}(a) shows that, in the absence of SOC, the BiSb monolayer behaves as a direct bandgap semiconductor with spin-degenerate electronic bands. With inclusion of SOC, the spin degeneracy is lifted, and a distinct Rashba spin splitting appears at the conduction band minima, with a Rashba parameter of \(\alpha_R = 1.9 \, \textrm{eV} \, \text{\AA}\), arising from the broken inversion symmetry [Fig.~\ref{fig:band structure}(b)]. In contrast, for the bilayer configuration, the restored inversion symmetry enforces the electronic bands to remain doubly degenerate, even in the presence of SOC. Consequently, no global Rashba spin splitting is observed in the bilayer, although each monolayer retains local RSOC -- a feature also reported in some other centrosymmetric systems~\cite{Yuan_Nat_comm_2019}.
Notably, while the monolayer exhibits semiconducting behavior, the bilayer displays a semi-metallic character with an electron pocket at the \(\Gamma\) point.

To understand the nature of the bands contributing to the electron pocket at the Fermi level (E$_F$), we compute the orbital-projected band structures of the optimized bilayer, with projections from the \(p_x + p_y\) and \(p_z\) orbitals of Bi and Sb atoms, as shown in Figs.~\ref{fig:band structure}(g) and \ref{fig:band structure}(h). 
We observe that the bands near the Fermi level are predominantly contributed by the Bi-\(p_z\) orbitals, while the Sb-\(p_z\) orbitals lie farther from E$_F$. 
This can be attributed to the fact that, as schematically illustrated in Fig.~\ref{fig:crystal structure}(b), the $p_z$ orbitals of the two Sb atoms in adjacent layers are in close proximity, resulting in strong repulsion between them. 
This steric repulsion pushes the Sb-\(p_z\)  states away from the Fermi level.  
In contrast, in the monolayer, the conduction band minimum is primarily contributed by the Bi-\(p_z\) and Sb-\(p_z\) orbitals, while the valence band maximum is mainly composed of Bi and Sb \(p_x + p_y\) orbitals, as shown in Fig.~\ref{monolayer} (Appendix~\ref{appendix c}).

The BiSb-SbBi bilayer maintains its semi-metallic behavior when two monolayers are brought closer together (i.e., \(d < d_0\)). However, as the interlayer distance increases beyond a critical value ($d_c=2.84$\,\AA), the system gradually transitions into a semiconducting phase (as shown in Fig.~\ref{band_evolution} in Appendix~\ref{appendix c}). This transition is accompanied by a concurrent shift from a topologically nontrivial ($Z_2=1$) to a trivial ($Z_2=0$) phase. 
When the separation between the two monolayers becomes sufficiently large, the interaction between the monolayers becomes minimal, and the electronic band structure of the bilayer resembles that of the free-standing monolayer configuration, as can be seen in Figs.~\ref{fig:band structure}(e) and \ref{fig:band structure}(f).

To gain deeper insight into the topological phase transitions in the bilayer configuration, we examine the roles of interlayer electron tunneling and SOC strength. 
When two BiSb monolayers are stacked in an inverted configuration, they effectively act as time-reversal conjugate partners. 
For smaller interlayer distances, i.e., below the critical value (\(d_c = 2.84 \, \text{\AA}\)), there is a finite probability of quantum tunneling of electrons between the constituent monolayers. 
When an electron hops from one monolayer to the other, it effectively switches with its time-reversal partner, resulting in a Z$_2$ invariant of 1 and the emergence of a topologically nontrivial phase. 
However, as the interlayer distance exceeds the critical value (\(d > d_c\)), electron tunneling is suppressed, causing the bilayer to transition into a topologically trivial phase with a Z$_2$ invariant of 0. 
Remarkably, even beyond this critical separation, the system can be driven back into a topologically nontrivial phase by artificially enhancing the SOC strength in the monolayers, as shown below. 

\begin{figure}[!!b]
\centering
\includegraphics[width=9cm, trim=19 0 0 0, clip]{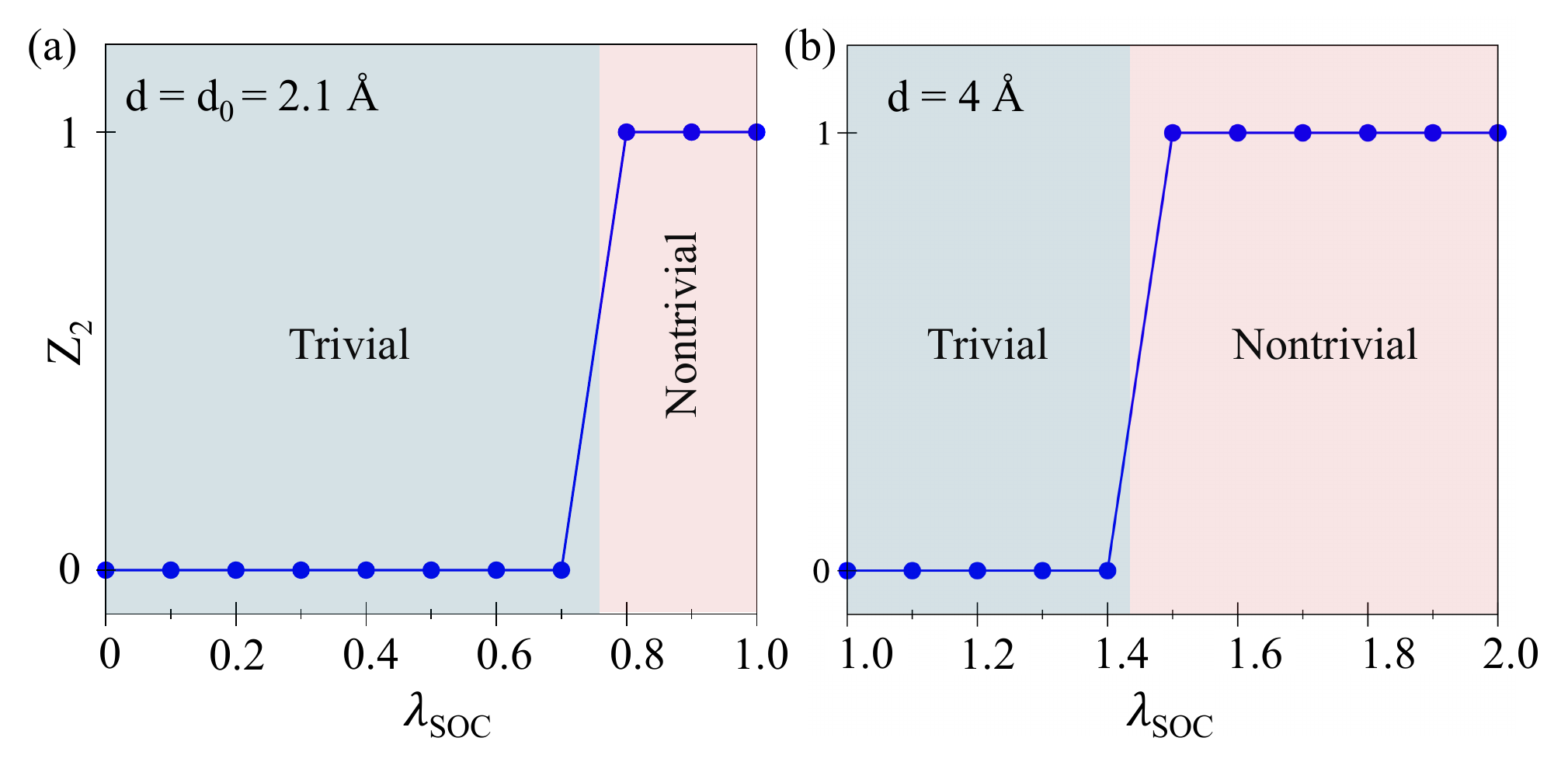}
\caption{The variation of the Z$_2$ topological invariant as a function of artificial SOC strength (\(\lambda_\textrm{SOC}\)) is shown for (a) \(d = 2.1 \, \text{Å}\) and (b) \(d = 4 \, \text{Å}\).}
\label{fig:tunable_SOC}
\end{figure}

{\bf Role of SOC strength on topology.}
The impact of SOC strength can be studied by artificially tuning the effective atomic SOC strength, \(\lambda_\textrm{SOC}\). 
In Fig.~\ref{fig:tunable_SOC}, we plot the Z$_2$ invariant as a function of \(\lambda_\textrm{SOC}\) for two limiting cases ($d<d_c$ and $d>d_c$), where \(\lambda_\textrm{SOC} = 1\) corresponds to the bare atomic SOC strength in our first-principles DFT calculations. 
In the topologically nontrivial case ($d < d_c$) with an optimized interlayer distance, a reduction of $\lambda_{\text{SOC}}$ by 30\% is required to drive the system into a topologically trivial phase [Fig.~\ref{fig:tunable_SOC}(a)] 
Conversely, in the topologically trivial case with $d > d_c$, an enhancement of $\lambda_{\text{SOC}}$ by more than 40\% is needed to induce a topologically nontrivial phase [Fig.~\ref{fig:tunable_SOC}(b)]. 

The evolution of the topological properties with respect to \(d\) and \(\lambda_\textrm{SOC}\) emphasizes that neither SOC nor interlayer electron tunneling (\(t\)) alone dictates the topological nature of the bilayer. 
Rather, it is the subtle balance between SOC and tunneling that governs the system's topology. 
Inspired from Ref.~\cite{Kaplan2024}, we find that this observation aligns with the scaling of the effective $g$-factor as \(\lambda_{SOC}^2/t\), as discussed below.

\begin{figure}[!t]
\centering
\includegraphics [width=1\columnwidth]{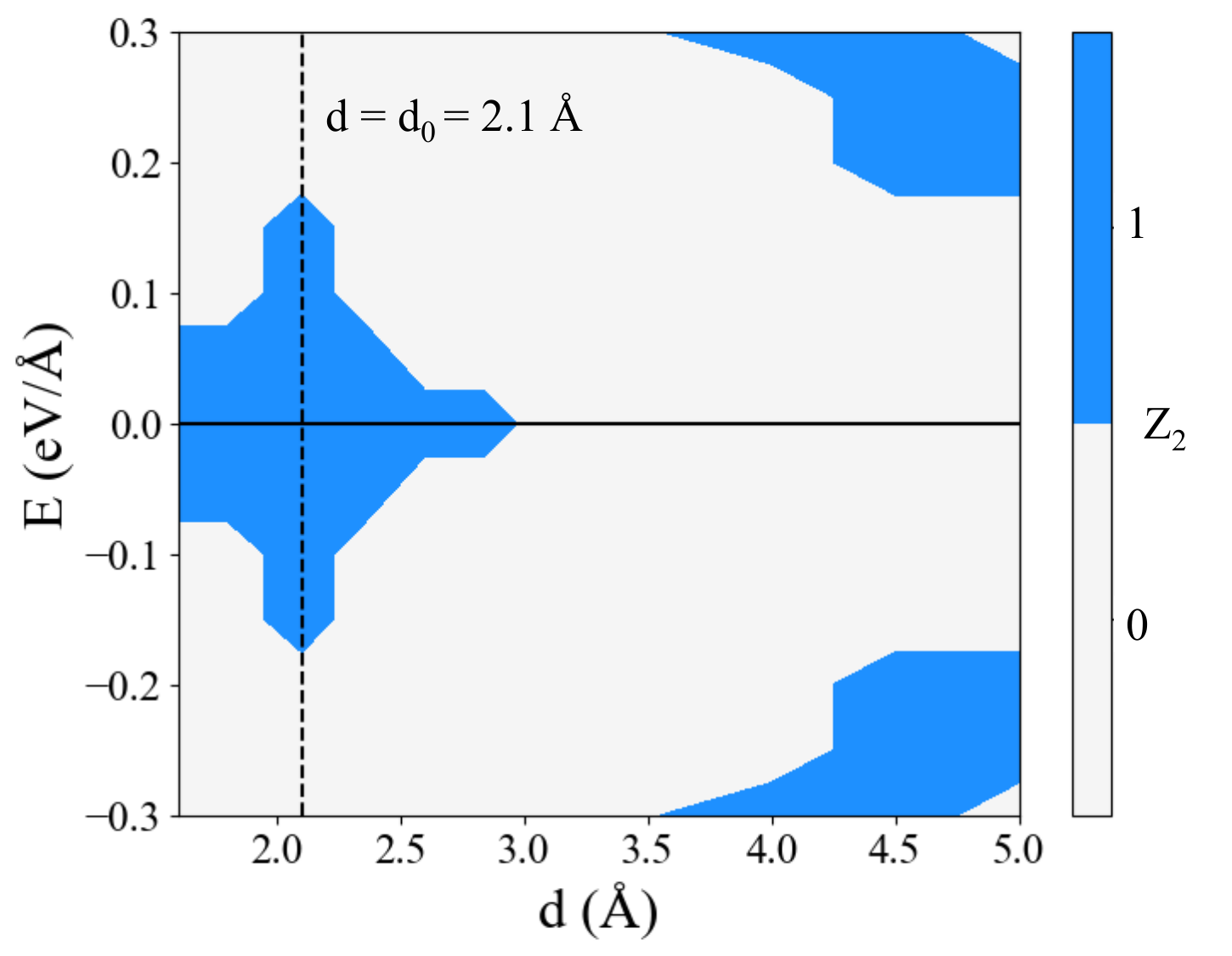}
\caption{2D phase diagram illustrating the variation of the Z$_2$ topological invariant in the BiSb-sbbi bilayer as a function of interlayer distance (\(d\)) and applied electric field (\(E\)). The color bar represents the corresponding values of the Z$_2$ invariant.}
\label{fig:phase diagram}
\end{figure}

We further predict that an experimentally tunable knob -- a perpendicular electric field $E$ -- enables the bilayer to toggle between topologically nontrivial and trivial phases. Fig.~\ref{fig:phase diagram} illustrates a 2D phase diagram showing the variation of the Z$_2$ topological invariant as a function of $d$ and $E$. 
The optimized bilayer (\(d = d_0\,=\,2.1\, \text{\AA}\)), which is topologically nontrivial in the absence of an electric field, retains this nontrivial character up to an applied field of $\sim$\(0.15\, \text{eV/\AA}\). Beyond this threshold, it transitions to a Z$_2$\,=\,0 phase.  In contrast, for the case of \(d=5\,\text{\AA}\), the system is initially topologically trivial (Z$_2$ = 0) at zero electric field. However, applying an electric field of around \(0.2\,\text{eV/\AA}\) induces a transition to a topologically nontrivial phase with Z$_2$ = 1. This reveals that the external electric field effectively modulates interlayer electron tunneling, providing a powerful knob for switching the system's topological phase.

\subsection*{Continuum modeling and magnetotransport}

To complete the description of the topological transition,  we examine its implications for experimentally accessible magnetotransport properties. 
In general, this problem involves the construction of Landau levels (LLs) for periodic systems within the space of localized orbitals, e.g., maximally localized Wannier functions~\cite{Marzari_RevModPhys.84.1419}, and hoppings dressed by the Peirels substitution \cite{Lado_2016}. 
In the limit of an infinite crystal, this would lead to the formation of a Hofstatder spectrum \cite{rossler2007bloch} and the breakdown of Bloch states for magnetic fields that are not rationally connected to the periodicity of the lattice \cite{azbel1964energy,Hofstadter1976spec}. 
However, since the mean-free path of electrons in Bi-Sb alloys \cite{white1958thermal,Jain1959} is not expected to exceed $\lambda \sim a_0$ (where $a_0$ is the lattice constant) coherence effects can be safely neglected.
As a result, Landau levels can be reasonably described using a quasi-classical expansion around the Fermi level $E_F$. 
In the Wannier basis, the Bloch Hamiltonian reads,
\begin{align}
    H_{\alpha \beta}(\mathbf{k}) = \sum_{\mathbf{R}} t_{\alpha \beta}(\mathbf{R}_i)e^{i\mathbf{k}\mathbf{R}_i},
    \label{eq:wan_ham}
\end{align}
such that $R_i = na_1 +ma_2$ are the unit cells, $t_{\alpha \beta}$ is the hopping matrix in the Wannier basis, and the total energy of the system is $H = \sum_{\alpha\beta,\mathbf{k}}H_{\alpha\beta}(\mathbf{k})c^\dagger_{\alpha \mathbf{k}}c_{\beta \mathbf{k}}$. $\alpha \beta$ are the orbital indices which include spin and $c^\dagger_\mathbf{\alpha k}$ is its associated creation operator. Expanding near the $\Gamma$ pocket where the largest density of electrons is present (Fig.~\ref{fig:band structure}), 
\begin{align}
    H_{\alpha \beta}(\mathbf{k}) = \sum_n w^n_{\alpha \beta} \mathbf{k}^n,
    \label{eq:continuum_wan}
\end{align}
where $w^n_{\alpha \beta}(\mathbf{k}) = \frac{(i)^n }{n!}\sum_{\mathbf{R}_i} t_{\alpha \beta} (\mathbf{R}_i) ( \hat{k}\cdot \mathbf{R}_i)\mathbf{R}_i^n$. At zeroth order in the expansion, we obtain the chemical potential for each orbital. The first order correction $w^1$ gives the linear spin-orbit coupling contributions in the form of Rashba and Dresselhaus effects, when projected onto the Bloch basis. $w_2$ encodes the effective mass, while $w_3$ includes higher-order SOC and trigonal warping effects~\cite{Fu2009}. As stated above, we apply a multiorbital quasiclassical quantization \cite{azbel1964energy}. Working in the Landau gauge $\mathbf{A} = (0,Bx,0)$, $\mathbf{B}=(0,0,B)$ we have,
\begin{align}
     k_x = \frac{i}{\sqrt{2} l_B}(a-a^\dagger), ~ k_y = \frac{1}{\sqrt{2}l_B}(a+a^\dagger),
\end{align}
such that $a,a^\dagger$ are bosonic ladder operators satisfying $[a,a^\dagger]=1$. $l_B = \sqrt{\frac{\hbar}{e B}}$ is the magnetic length. This choice correctly preserves the canonical commutation relations in a magnetic field for free particles, $i\hbar^2 [k_x,k_y] = -ie\hbar B$ \cite{tong2016lectures}. Constructing the LL basis from the vacuum $a |0\rangle=0$, allows us to recast the problem in the LL basis and obtain an eigenequation,
\begin{align}
    (H_{\alpha\beta}^{l_1,l_2}+g_0B\sigma_z^{\alpha\beta} \delta_{l_1,l_2}) \psi_{l_1,l_2,\alpha\beta} = \varepsilon \psi_{l_1,l_2,\alpha\beta},
\end{align}
where $l_1,l_2$ are LL indices. In addition, we added Zeeman coupling for the spin components of the $\alpha\beta$ indices. The coefficients of $H$ are obtained by projecting the quantized momentum operators $k_x,k_y$ onto the LL basis. e.g., $\langle l_1|k_x|l_2\rangle =\frac{i}{\sqrt{2} l_B}\left(\sqrt{l_2}\delta_{l_1,l_2-1}-\sqrt{l_2+1}\delta_{l_1,l_2+1}\right)$, and $\langle l_1|k_y|l_2\rangle = \frac{1}{\sqrt{2} l_B}\left(\sqrt{l_2}\delta_{l_1,l_2-1}+\sqrt{l_2+1}\delta_{l_1,l_2+1}\right)$. The continuum model and LLs are constructed in the following manner. For a basis of $N_w$ Wannier functions (including spin), the Hamiltonian takes the generic form,
\begin{align}
    H=\left(
    \begin{matrix}
        H_{N_w} & H_{01} & 0 & \ldots \\
        H_{01} & H_{N_w}  & H_{21} & \ldots\\ 
            0  &     H_{12} & H_{N_w} & \ldots
    \end{matrix}\right),
\end{align}
where $H_{N_w}$ is the continuum expansion of the Wannier Hamiltonian in Eq.~\eqref{eq:continuum_wan}. Every block of $H$ represents a different Landau level index, $n=0,1,2,\ldots, N_L$. Every block is coupled to neighboring blocks with the terms $H_{lm}$. In the calculation we keep $N_L = 20$ LLs. Thus, $H$ is of the size $N_w^2 N_L$. We have verified this value of $N_L$ is convergent w.r.t the LL spectrum.

\begin{figure}[!!t]
\centering
\includegraphics [width=1\columnwidth]{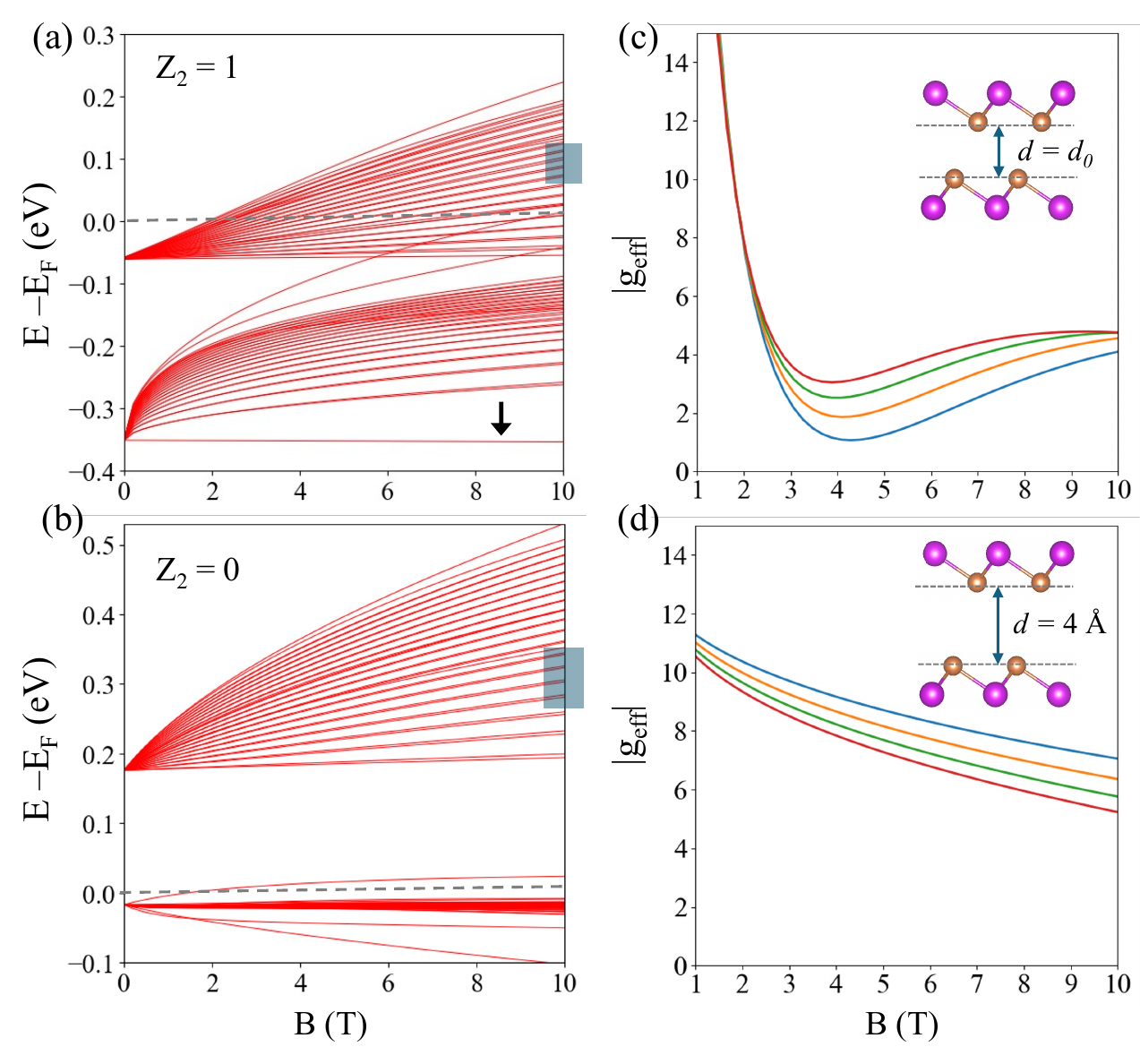}
\caption{Landau levels computed for the BiSb-SbBi bilayer at (a) \(d = 2.1 \, \text{Å}\) (topologically nontrivial case) and (b) \(d = 4 \, \text{Å}\) (topologically trivial case).  
The variation of the effective \(g\)-factor is presented for (c) \(d = 2.1 \, \text{Å}\) and (d) \(d = 4 \, \text{Å}\). The effective \(g\)-factors are computed for the Landau levels highlighted by the blue shaded regions in (a) and (b).
}
\label{fig:Hall conductivity and g factor}
\end{figure}

\begin{figure}[!!b]
\centering
\includegraphics [width=9.8cm]{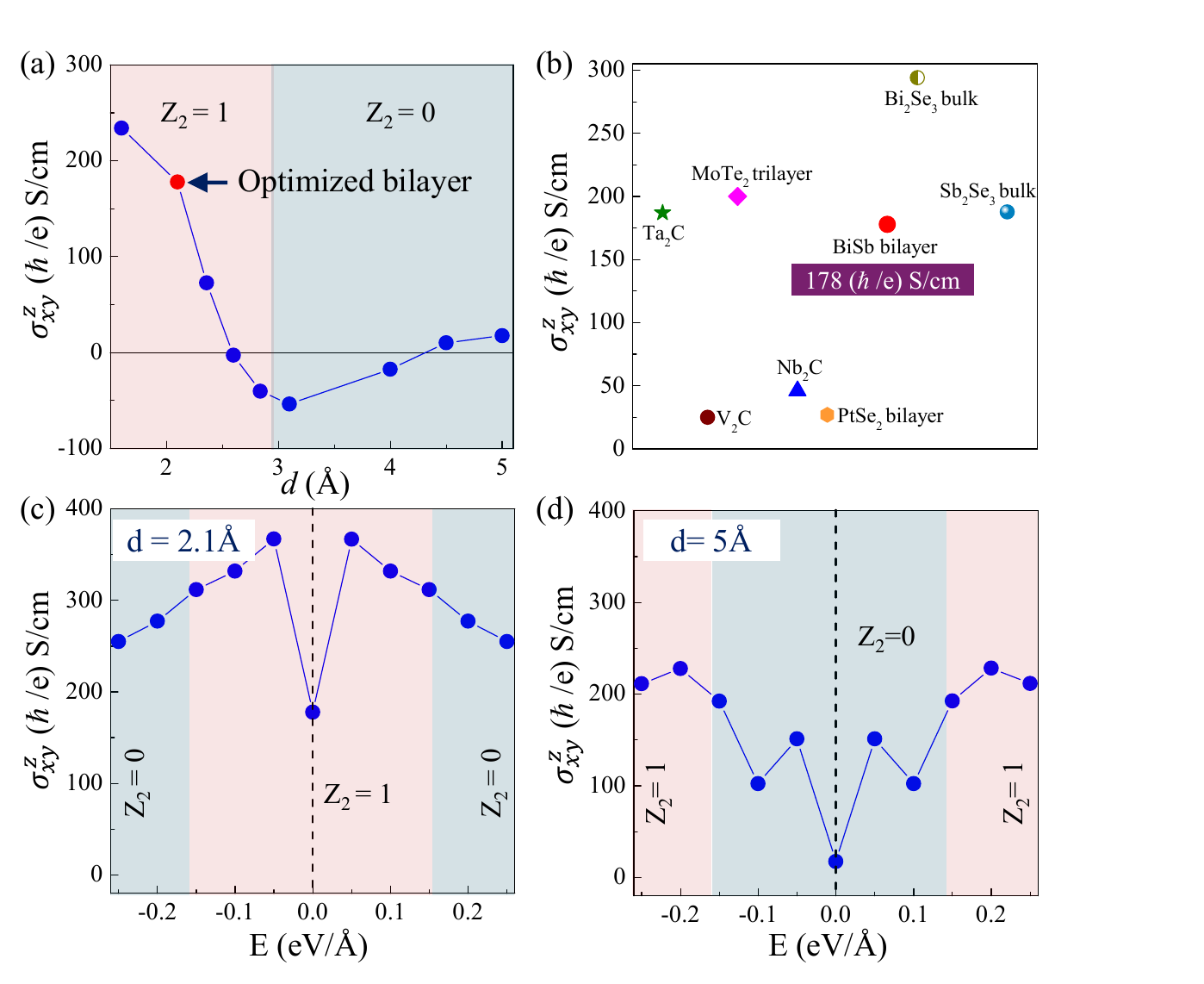}
\caption{Variation of the transverse component of spin Hall conductivity (\(\sigma_{xy}^z\)) in the BiSb-SbBi bilayer. (a) \(\sigma_{xy}^z\) at the Fermi level as a function of interlayer distance (\(d\)). (b) Comparison of \(\sigma_{xy}^z\) at the Fermi level for the bilayer with \(d = 2.1 \, \text{Å}\) and other 2D materials, including M$_2$C monolayers (M = Nb, Ta, V)~\cite{Zuo_PhysRevB_2023}, PtSe$_2$ bilayer~\cite{Li_PhysRevB_2021}, MoTe$_2$ trilayer~\cite{Song_Nat_mat_2020}, and well-known 3D topological insulators such as Bi$_2$Se$_3$ and Sb$_2$Se$_3$~\cite{Farzaneh_PhysRevMaterials.4.114202}. (c) \(\sigma_{xy}^z\) at the Fermi level for the bilayer with \(d = 2.1 \, \text{Å}\) under an applied electric field. (d) \(\sigma_{xy}^z\) at the Fermi level for the bilayer with \(d = 5 \, \text{Å}\) under an applied electric field. The light red region indicates the topologically nontrivial phase with Z$_2$ = 1, while the light blue region represents the topologically trivial phase with Z$_2$ = 0.}
\label{fig:SHC}
\end{figure}

The resultant LL spectrum ($\varepsilon_{\tilde{n}}$) is plotted in Fig.~\ref{fig:Hall conductivity and g factor}(a-b). For the nontrivial $Z_2=1$ phase [Fig.~\ref{fig:Hall conductivity and g factor}(a)], we find strongly dispersive LLs, but with a very weakly dispersive lowest-LL (near the bottom of the band at $\approx 0.35\,\textrm{eV}$ below E$_F$), as marked in Fig.~\ref{fig:Hall conductivity and g factor}(a) using a black arrow. 
This is a direct indication of the presence of a non-trivial surface state \cite{Lado_2016}. 
Following Ref.~\cite{Lado_2016}, we find its Dirac-index $\gamma \approx -0.005$ suggesting its origin is from a Dirac-like surface state, consistent with the non-trivial phase. No such state is observed for $Z_2 = 0$ [Fig.~\ref{fig:Hall conductivity and g factor}(b)] where bands are curved consistent with the picture of spin-orbit split Landau levels \cite{winkler2003spin,Kaplan2024}.

Furthermore, band mixing in the trivial state is significantly stronger as evidenced by the curvature of Landau levels. In addition, the sign of the LL dispersion, in the regime of $B\to 0$ is indicative of a band inversion. In general \cite{Kaplan2024,winkler2003spin}, the LL for small magnetic field is strongly altered by SOC, $E_{n\sigma} \approx \sigma\sqrt{n}\alpha l_B, B\to 0$. Here $\sigma = \pm 1$ is the pseudospin index of the LL. 
For a simple parabolic band with spin \cite{Wang2003}, the LLs split into two sectors that disperse with opposite (in sign) slope with $B$. 
In the topological case, i.e., Fig.~\ref{fig:Hall conductivity and g factor}(a), all Landau levels, with the exception of the zeroth one, disperse with a positive slope, indicating the strong renormalization of SOC due to electron tunneling. 
This contrasts with Fig.~\ref{fig:Hall conductivity and g factor}(b), which shows two branches of the LLs dispersing with opposite signs, as normally expected~\cite{Kaplan2024}. Thus, the Landau fan itself marks the topological transition.

The strength of SOC has immediate effect on the calculated effective $g$-factors of the LLs and the energy splittings. However, the $g$ factors are also affected by the interlayer hybridization~\cite{Kaplan2024}. 
In Figs.~\ref{fig:Hall conductivity and g factor}(c)-(d) we compare the $g$-factors of LLs in the vicinity of $E_F$ for the topological (c) and trivial (d) states. The suppression of the $g$-factor, which indicates stronger tunneling between layers, can be clearly observed in the topological (c) case. Following the approach of Kaplan \textit{et al.}~\cite{Kaplan2024}, the effective $g$-factor can be estimated by $g_\textrm{eff} \sim \lambda_{SOC}^2/\mu_B t$, which directly demonstrates its dependence on the interlayer tunneling $t$.

\subsection*{Tunable spin Hall effect}
The presence of a nontrivial topological phase transition, combined with strong local RSOC in the constituent monolayers, underscores the potential for a significant spin Hall effect (SHE) in the studied bilayers~\cite{Hirsch_PhysRevLett.83.1834, DYAKONOV_phy_let_a_1971459, Jungwirth_Nature_Mat_2012, Sinova_PRL_2004, Gong_PhysRevB.109.045124, Bordoloi_PhysRevB_2024}. 

According to linear response theory, the spin Hall conductivity (SHC) represented as \({\sigma_{ij}^k}\) relates the applied electric field \(E_j\) to the spin current density \({J_{i}^k}\) as \({J_{i}^k} = {\sigma_{ij}^k} E_j\). Here, the indices \(i\), \(j\), and \(k\) represent mutually perpendicular directions. While SHC is generally a third-rank tensor with 27 components, symmetry constraints for a system with $p\bar{3}m1$ layer group symmetry allow only six nonzero components: \(\sigma_{xx}^x\), \(\sigma_{yy}^x\), \(\sigma_{xy}^y\), \(\sigma_{yx}^y\), \(\sigma_{xy}^z\), and \(\sigma_{yx}^z\)~\cite{Aroyo_Acta_Crystallographica_Section_A_2006,Aroyo_Bulgarian_Chemical_Comm_2011}. Furthermore, these components reduce to only two independent terms due to symmetry relations: \(\sigma_{xx}^x\,=\,-\sigma_{yy}^x\,=\,-\sigma_{xy}^y\,=\,-\sigma_{yx}^y\) and \(\sigma_{xy}^z\,=\,-\sigma_{yx}^z\).

Due to the finite bandgap in the topologically trivial BiSb monolayer, SHC near $E_F$ is negligible. However, with the emergence of the topological phase in the BiSb bilayer, the transverse component of the SHC (\(\sigma_{xy}^z\)) increases significantly, reaching approximately 178\,\((\hbar/e) \, \text{S/cm}^{-1}\) at $E_F$, as illustrated in Fig.~\ref{fig:SHC}(a). A comparative analysis presented in Fig.~\ref{fig:SHC}(b) shows that the SHC value for the BiSb-SbBi bilayer surpasses those of many other 2D TIs and is comparable to some well-known bulk TIs such as Sb\(_2\)Se\(_3\)~\cite{Farzaneh_PhysRevMaterials.4.114202}.

Similar to previously studied spin Hall effects \cite{kim2012topological,collins2018electric}, SHC of the bilayer can also be tuned by varying $d$ and a vertical displacement field $E$. Fig.~\ref{fig:SHC}(a) illustrates the dependence of \(\sigma_{xy}^z\) at $E_F$ on $d$, while Figs. \ref{fig:SHC}(c) and (d) show the SHC dependence on $E$ for the optimized bilayer and the bilayer at \(d = 5\,\text{\AA}\) respectively. With decreasing $d$, i.e, bringing the layers closer together than the optimized interlayer distance enhances \(\sigma_{xy}^z\) at \(E_F\) to approximately 234\,\((\hbar/e)\,\text{S/cm}^{-1}\). On the other hand, as $d$ increases, \(\sigma_{xy}^z\) initially decreases but increases again beyond a critical separation. Strikingly, with an applied electric field, \(\sigma_{xy}^z\) for the optimized bilayer increases from 178 to approximately 367\,\((\hbar/e) \, \text{S/cm}^{-1}\) under a 0.05 eV/$\AA$ field. With further increase in $E$, \(\sigma_{xy}^z\) gradually decreases but still remains higher than its zero-field value. Furthermore, for \(d = 5 \, \text{\AA}\), \(\sigma_{xy}^z\) increases with the electric field but exhibits a non-monotonic trend. This tunability of SHC via electric fields offers a promising method for controlling SHC in practical applications, emphasizing the potential of BiSb bilayers for next-generation spintronic devices.

\section*{Conclusions}
In this work, we demonstrated that stacking two topologically trivial BiSb monolayers ($Z_2=0$)  in an inverted BiSb-SbBi bilayer configuration induces a nontrivial  topological phase ($Z_2=1$), governed by the interplay between SOC and interlayer distance. 
While the monolayer exhibits strong SOC (${\alpha}_R = 1.9\,\mathrm{eV} \cdot \mathrm{\AA}$), it remains topologically trivial in isolation; topological order emerges only through inverted bilayer stacking. 
By systematically exploring the parameter space, we showed that neither SOC strength nor interlayer spacing alone dictates the topological transition -- instead, it is governed by the ratio $\lambda_{SOC}^2/t$, where $\lambda_{SOC}$ is the SOC strength and $t$ is the kinetic tunneling parameter along the interlayer vdW direction. 
Following Ref.~\cite{Kaplan2024}, we argue that the ratio $\lambda_{\text{SOC}}^2/t$ governs the Landau level spectrum, with the suppression of the effective $g$-factor. 
Crucially, we identify the emergence of a zeroth Landau level, offering a clear experimental signature of nontrivial topology even in metallic regimes. 
These results suggest that $\lambda_{\text{SOC}}^2/t$ can serve as an effective target parameter in materials searches for topological states in multilayer vdW heterostructures. 

The strength of SOC -- and its connection to topology -- can be experimentally probed in this system by applying an out-of-plane electric field. This field breaks structural inversion symmetry and modulates the Rashba parameter, enabling reversible switching between topologically nontrivial and trivial phases.

We propose utilizing the topological properties of materials assembled via our approach for spintronic applications~\cite{Zutic_2004_spintronics}. 
The bilayer structure exhibits a substantial, gate-tunable SHC of $\sim$178\,($\hbar/e$)\,S/cm at the Fermi level, exceeding that of many 2D TIs and rivaling some bulk TIs. 
Notably, the spin Hall effect is highly sensitive to an applied out-of-plane electric field, suggesting the feasibility of spin-transistor functionalities, where spin-dependent properties can be efficiently controlled via gate voltage~\cite{Zutic_2004_spintronics}.


A distinct new avenue for topological switching that we envisage involves light–matter interaction, in a similar manner to a recent proposal for the control of ferroic order in bilayers~\cite{wei2025ultrafast}. In addition, spatial order and topological domains could be realized by coupling light to the out-of-plane phonon mode that modulates the interlayer vdW distance \cite{Kaplan2025_spatio}. 
These unique features, combined with exciting possibilities of realizing exotic quantum phases by incorporating superconducting or ferromagnetic components into the multilayer matrix, make our approach promising for next-generation spintronic and hybrid materials applications. 

\section*{Acknowledgements}
AB and SS acknowledge support from the U.S.~Department of Energy, Office of Science, Office of Fusion Energy Sciences, Quantum Information Science program under Award No.~DE-SC-0020340.
DK is supported by an Abrahams postdoctoral fellowship of the Center for Materials Theory, Rutgers University and the Zuckerman STEM fellowship. DK thanks the hospitality of the Aspen Center for Physics, which is supported by National Science Foundation grant PHY-2210452. DK also acknowledges partial support by grant NSF PHY-2309135 to the Kavli Institute for Theoretical Physics (KITP) where part of this work was finalized.
Authors thank the Pittsburgh Supercomputer Center (Bridges2) supported by the Advanced Cyberinfrastructure Coordination Ecosystem: Services \& Support (ACCESS) program, which is supported by National Science Foundation grants \#2138259, \#2138286, \#2138307, \#2137603, and \#2138296.  

\clearpage
\appendix
\section{Phonon band structure}
\begin{figure}[tbh]
\centering
\includegraphics[trim=0cm 8cm 0cm 0cm, clip, width=1.0\columnwidth]{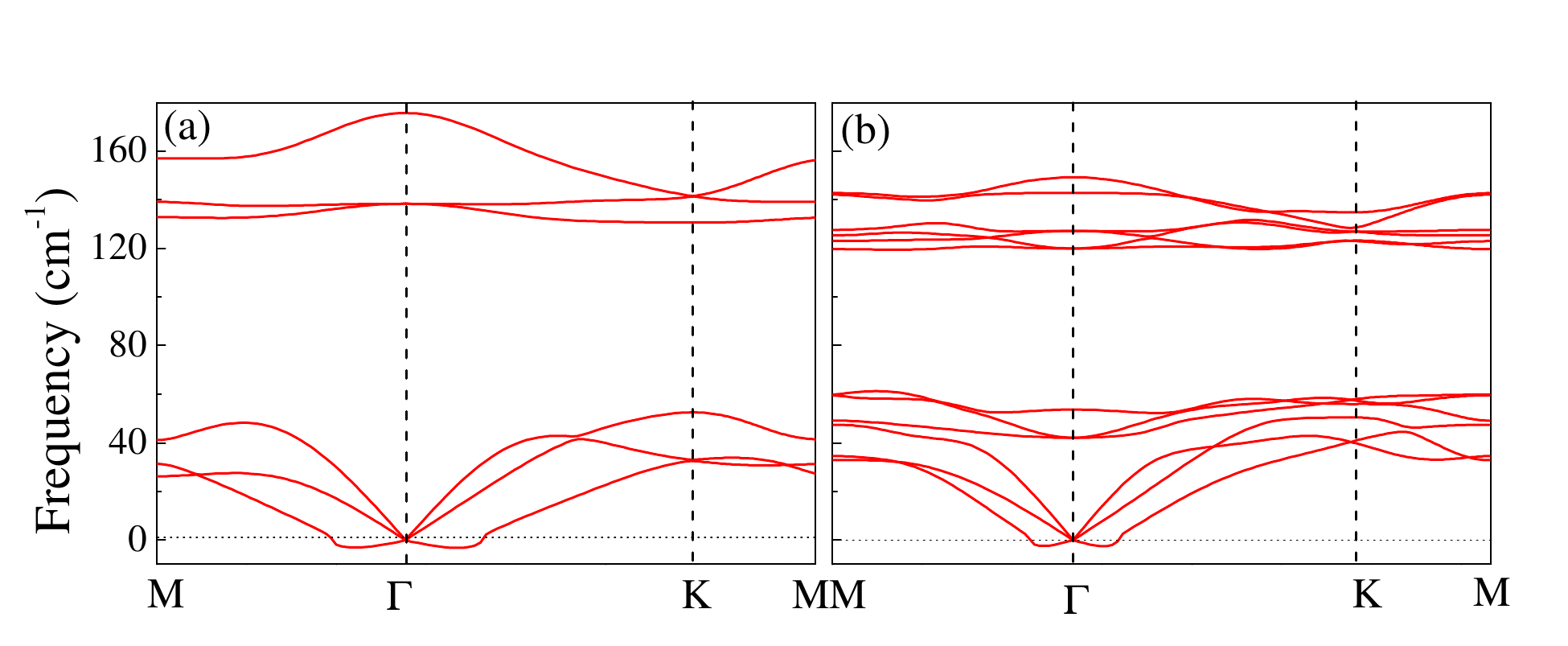}

\caption{Phonon bandstructure of (a) BiSb monolayer and (b) BiSb–SbBi bilayer, computed along the high-symmetry directions in the hexagonal Brillouin zone.}
\label{phonon}
\end{figure}

Figure~\ref{phonon} shows the phonon band structures of (a) the optimized BiSb monolayer and (b) the BiSb–SbBi bilayer configurations, computed along the high-symmetry directions of the Brillouin zone. Notably, no imaginary phonon modes are present in the phonon band structures, except for small U-shaped features observed in the first acoustic branch (ZA) near the $\Gamma$ point. However, these U-shaped features do not signify any lattice instability; rather, they correspond to poor numerical convergence near zone center ($q\rightarrow0$) and the flexural acoustic modes commonly present in 2D and layered van der Waals materials~\cite{Stengel_PhysRevX.11.041027, Singh_PhysRevB_2017, SinghPRL2020}.

\section{Raman spectra}
\label{appendix B}
\begin{figure}[tbh]
\centering
\includegraphics[trim=0cm 0cm 0cm 0cm, clip, width=1.0\columnwidth]{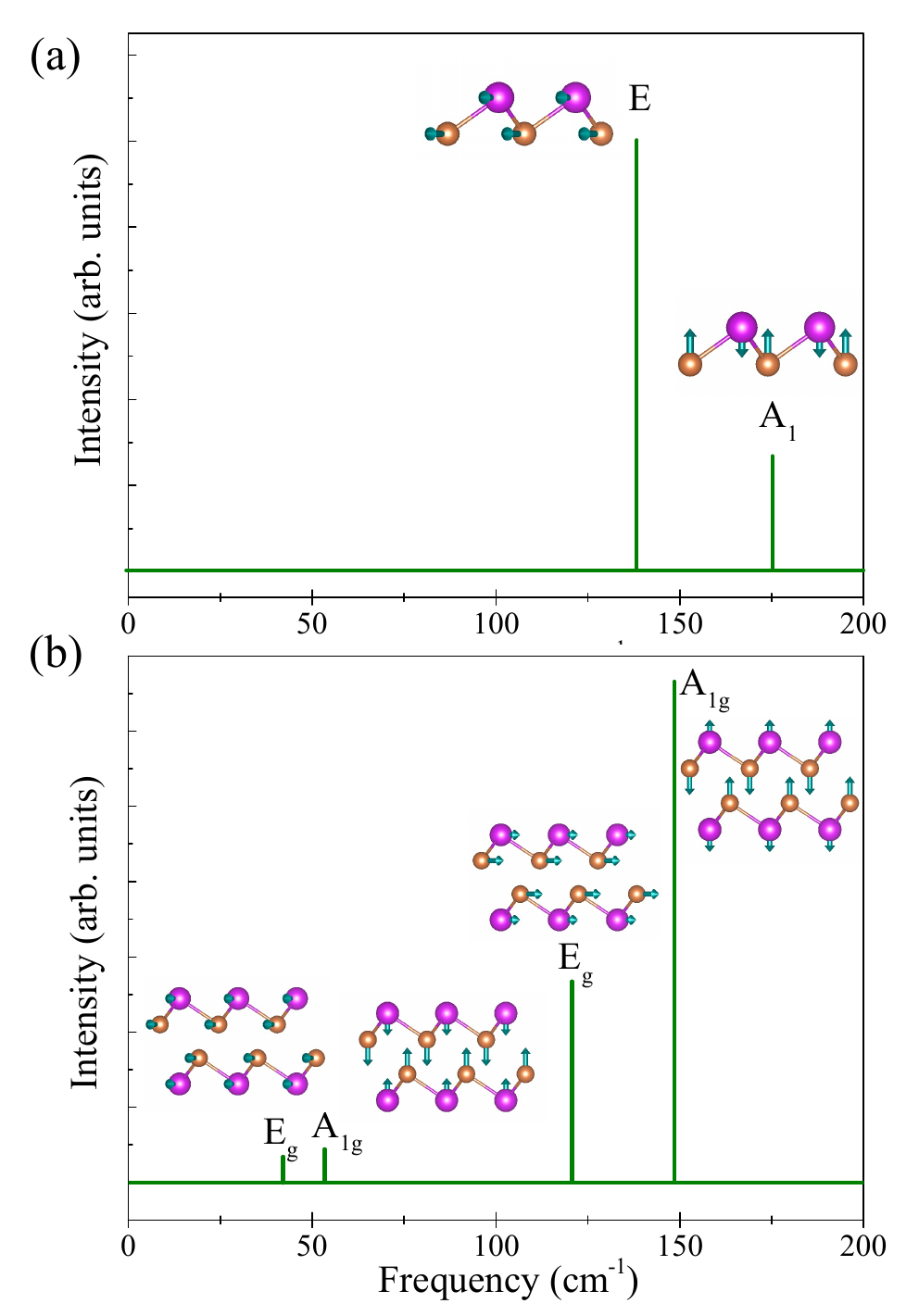}

\caption{Computed Raman spectra of (a) BiSb monolayer and (b) BiSb-SbBi bilayer. Insets show the atomic displacement patterns associated with each Raman-active mode.}
\label{Raman}
\end{figure}

Figure \ref{Raman}(a) shows the calculated Raman spectra corresponding to the BiSb monolayer, revealing two distinct Raman-active modes: a doubly degenerate E mode at 138.4 cm$^{-1}$, and a non-degenerate A$_1$ mode at 175.7 cm$^{-1}$. 
These modes are both Raman and infrared (IR) active.
In contrast, the simulated Raman spectrum for the centrosymmetric BiSb-SbBi bilayer features four distinct Raman-active modes: 
two doubly degenerate E$_{g}$ modes at 42.1 and 120.0\,cm$^{-1}$ frequencies, and two non-degenerate A$_{1g}$ modes  
at 53.7 and 149.2\,cm$^{-1}$ [Figure \ref{Raman}(b)]. 
The atomic displacement patterns corresponding to these modes are shown in Figure~\ref{Raman}.
The distinctive features present in the simulated Raman spectra can be used to differentiate between the monolayer and bilayer configurations in future experiments.

\section{Electronic band structure}
\label{appendix c}
\begin{figure}[!!b]
\centering
\includegraphics[trim=0.5cm 1cm 0cm 0cm, clip, width=9.2cm]{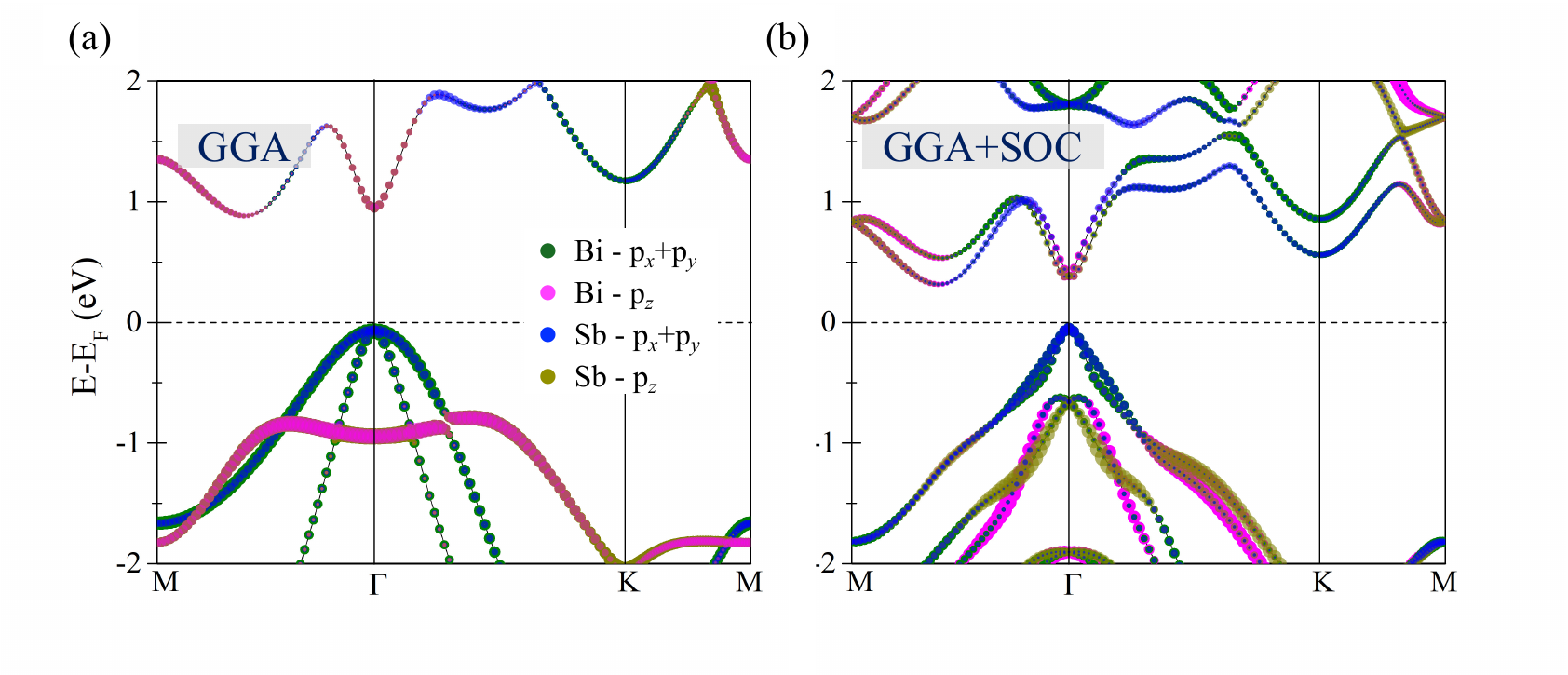}

\caption{Orbital-projected electronic band structure of the BiSb monolayer computed along the high-symmetry directions of the Brillouin zone (a) without SOC and (b) with SOC. The dashed line indicates the Fermi level.}
\label{monolayer}
\end{figure}

\begin{figure*}
\centering
\includegraphics[width=19cm]
{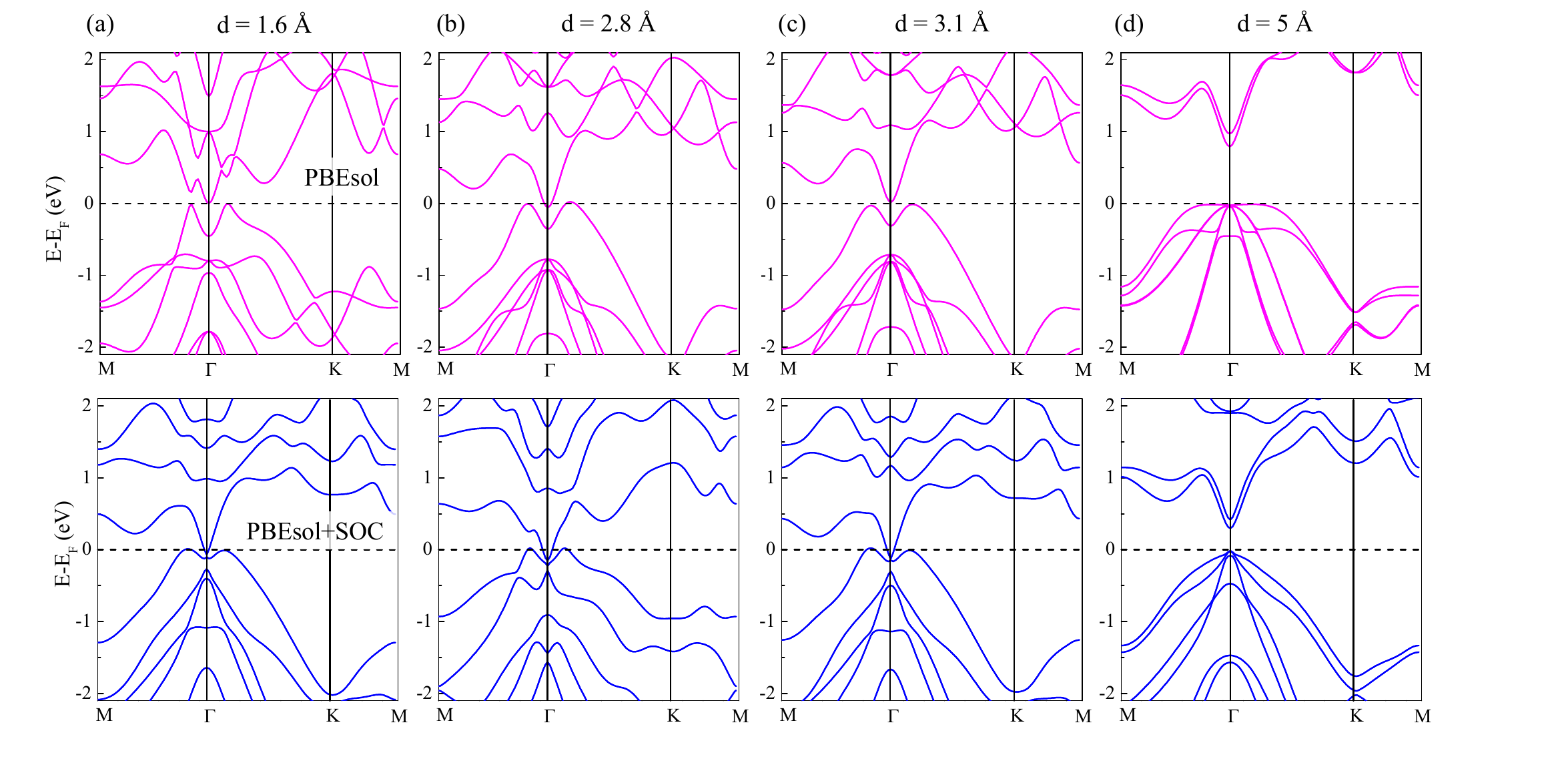}
\caption{Evolution of electronic band structures of the BiSb-SbBi bilayer with varying interlayer distance $d$. Electronic band structures computed along high-symmetry directions for (a) $d$=1.6 Å, (b) $d$=2.8 Å, (c) $d$=3.1 Å, and (d) $d$=5.0 Å. Horizontal dashed lines indicate the Fermi level. The band structures in the top panel are calculated without SOC (magenta), while those in the bottom panel are with calculated SOC (blue).}
\label{band_evolution}
\end{figure*}

Fig.~\ref{monolayer} shows the orbital-projected electronic band structure of the BiSb monolayer calculated (a) without and (b) with spin-orbit coupling (SOC). The conduction band minimum is primarily contributed by the Bi-\(p_z\) and Sb-\(p_z\) orbitals, while the valence band maximum is mainly composed of contributions from the Bi and Sb \(p_x + p_y\) orbitals.

\vspace{0.4 cm}
Figure \ref{band_evolution} represents the evolution of the electronic band structure in the bilayer while systematically varying the interlayer separation ($d$), keeping all other parameters constant. The figure presents electronic band structures for four representative values of $d$: 1.6, 2.8, 3.1, and 5\,\AA. From the band structures it is evident that when the two monolayers are brought into close proximity below $d_0=$ 2.1\,\AA, the bilayer maintains its semi-metallic character with a prominent electron pocket at the $\Gamma$ point. On the other hand, as the inter-layer distance exceeds $d_0 = 2.1$\,\AA\ and is increased upto a critical value ($d_c = 2.84$\,\AA), the system continues to exhibit its semi-metallic behaviour. Remarkably, beyond this critical value $d_c$, the system undergoes a transition into a semiconducting phase. Finally, as the separation between the two monolayers increases sufficiently, the electronic band structure tends to resemble that of the free-standing monolayer configuration.
\vspace{1cm}
\bibliography{biblio}
\end{document}